\overfullrule=0pt
\input harvmac
\lref\jusinskas{
   N.~Berkovits and R.~Lipinski Jusinskas,
  ``Light-Cone Analysis of the Pure Spinor Formalism for the Superstring,''
JHEP {\bf 1408}, 102 (2014).
[arXiv:1406.2290 [hep-th]].
}
\lref\AisakaUD{
  Y.~Aisaka, L.~I.~Bevilaqua and B.~C.~Vallilo,
  ``On semiclassical analysis of pure spinor superstring in an $AdS_5$ x $S^5$ background,''
JHEP {\bf 1209}, 068 (2012).
[arXiv:1206.5134 [hep-th]].
}

\lref\brink{
 L.~Brink, M.~B.~Green and J.~H.~Schwarz,
  ``Ten-dimensional Supersymmetric {Yang-Mills} Theory With SO(8) - Covariant Light Cone Superfields,''
Nucl.\ Phys.\ B {\bf 223}, 125 (1983)..
}

\lref\BerkovitsYR{
  N.~Berkovits and O.~Chandia,
  ``Superstring vertex operators in an AdS(5) x S**5 background'',
Nucl.\ Phys.\ B {\bf 596}, 185 (2001).
[hep-th/0009168].
}

\lref\BerkovitsRB{
  N.~Berkovits,
``Covariant quantization of the superparticle using pure spinors,''
JHEP {\bf 0109}, 016 (2001).
[hep-th/0105050].
}

\lref\GalperinAV{
  A.~Galperin, E.~Ivanov, S.~Kalitsyn, V.~Ogievetsky and E.~Sokatchev,
``Unconstrained N=2 Matter, Yang-Mills and Supergravity Theories in Harmonic Superspace,''
Class.\ Quant.\ Grav.\  {\bf 1}, 469 (1984)..
}

\lref\vallilo{
  B.~Vallilo and L.~Mazzucato,
  ``The Konishi Multilpet at Strong Coupling,''
JHEP {\bf 1112}, 029 (2011).
[arXiv:1102.1219 [hep-th]].
}

\lref\MikhailovAF{
  A.~Mikhailov,
 ``Finite dimensional vertex,''
JHEP {\bf 1112}, 005 (2011).
[arXiv:1105.2231 [hep-th]].
}

\lref\mikh{
  A.~Mikhailov and R.~Xu, ``BRST cohomology of the sum of two pure spinors,''
to appear.
}

\lref\minahan{
  J.~Minahan,
  ``Holographic three-point functions for short operators,''
JHEP {\bf 1207}, 187 (2012).
[arXiv:1206.3129 [hep-th]].
}

\lref\BerkovitsGA{
  N.~Berkovits,
  ``Simplifying and Extending the AdS(5) x S**5 Pure Spinor Formalism,''
JHEP {\bf 0909}, 051 (2009).
[arXiv:0812.5074 [hep-th]].}



\lref\BerkovitsBT{
  N.~Berkovits,
  ``Pure spinor formalism as an N=2 topological string'',
JHEP {\bf 0510}, 089 (2005).
[hep-th/0509120].}

\lref\BerkovitsXU{
  N.~Berkovits,
 ``Quantum consistency of the superstring in AdS(5) x S**5 background,''
JHEP {\bf 0503}, 041 (2005).
[hep-th/0411170].
}


\lref\BerkovitsFE{
  N.~Berkovits,
  ``Super Poincare covariant quantization of the superstring'',
JHEP {\bf 0004}, 018 (2000).
[hep-th/0001035].
}


\lref\MazzucatoJT{
  L.~Mazzucato,
  ``Superstrings in AdS'',
[arXiv:1104.2604 [hep-th]].
}


\lref\HeslopNP{
  P.~Heslop and P.~S.~Howe,
  ``Chiral superfields in IIB supergravity'',
Phys.\ Lett.\ B {\bf 502}, 259 (2001).
[hep-th/0008047].
}







\lref\SohniusWK{
  M.~F.~Sohnius,
  ``Bianchi Identities for Supersymmetric Gauge Theories,''
Nucl.\ Phys.\ B {\bf 136}, 461 (1978).
}

\lref\ArutyunovGA{
  G.~Arutyunov and S.~Frolov,
  ``Foundations of the $AdS_5 \, x \, S^5$ Superstring. Part I,''
J.\ Phys.\ A {\bf 42}, 254003 (2009).
[arXiv:0901.4937 [hep-th]].
}

\lref\MetsaevIT{
  R.~R.~Metsaev and A.~A.~Tseytlin,
  ``Type IIB superstring action in AdS(5) x S**5 background,''
Nucl.\ Phys.\ B {\bf 533}, 109 (1998).
[hep-th/9805028].
}



\lref\HoweSRA{
  P.~S.~Howe and P.~C.~West,
  ``The Complete N=2, D=10 Supergravity'',
Nucl.\ Phys.\ B {\bf 238}, 181 (1984).
}

\lref\BerkovitsULM{
  N.~Berkovits,
  ``Sketching a Proof of the Maldacena Conjecture at Small Radius,''
[arXiv:1903.08264 [hep-th]].
}
\lref\GomezSLA{
  H.~Gomez and C.~R.~Mafra,
  ``The closed-string 3-loop amplitude and S-duality,''
JHEP {\bf 1310}, 217 (2013).
[arXiv:1308.6567 [hep-th]].
}
\lref\FrolovAV{
  S.~Frolov and A.~A.~Tseytlin,
  ``Semiclassical quantization of rotating superstring in AdS(5) x S**5,''
JHEP {\bf 0206}, 007 (2002).
[hep-th/0204226].
}
\lref\ChoNFN{
  M.~Cho, S.~Collier and X.~Yin,
  ``Strings in Ramond-Ramond Backgrounds from the Neveu-Schwarz-Ramond Formalism,''
[arXiv:1811.00032 [hep-th]].
}
\lref\MinahanFH{
  J.~A.~Minahan,
  ``Holographic three-point functions for short operators,''
JHEP {\bf 1207}, 187 (2012).
[arXiv:1206.3129 [hep-th]].
}
\lref\MazzucatoJaT{
  L.~Mazzucato,
  ``Superstrings in AdS,''
Phys.\ Rept.\  {\bf 521}, 1 (2012).
[arXiv:1104.2604 [hep-th]].
}

\lref\BerkovitsPS{
  N.~Berkovits and T.~Fleury,
  ``Harmonic Superspace from the $AdS_5\times S^5$ Pure Spinor Formalism,''
JHEP {\bf 1303}, 022 (2013).
[arXiv:1212.3296 [hep-th]].
}
\lref\BerkovitsULM{
  N.~Berkovits,
  ``Sketching a Proof of the Maldacena Conjecture at Small Radius,''
[arXiv:1903.08264 [hep-th]].
}
\lref\BedoyaQZ{
  O.~A.~Bedoya, L.~I.~Bevilaqua, A.~Mikhailov and V.~O.~Rivelles,
  ``Notes on beta-deformations of the pure spinor superstring in AdS(5) x S(5),''
Nucl.\ Phys.\ B {\bf 848}, 155 (2011).
[arXiv:1005.0049 [hep-th]].
}

\def\a{{\alpha}}

\def\ad{{\dot a}}

\def\bd{{\dot \b}}
\def\l{{\lambda}}
\def\lb{{\overline\lambda}}

\def\lb{{\overline\lambda}}
\def\b{{\beta}}

\def\g{{\gamma}}

\def\d{{\delta}}
\def\e{{\epsilon}}
\def\s{{\sigma}}

\def\N{{\nabla}}

\def\O{{\Omega}}

\def\half{{1\over 2}}
\def\p{{\partial}}

\def\t{{\theta}}
\def\tb{{\overline\theta}}

\def\ad{{\dot a}}
\def\bd{{\dot b}}
\def\cd{{\dot c}}
\def\dd{{\dot d}}

\def\tb{{\overline\theta}}

\def\lb{{\overline{\lambda}}}

\Title{\vbox{\baselineskip12pt
\hbox{}}}
{{\vbox{\centerline{Half-BPS Vertex Operators}
\smallskip
\centerline{of the $AdS_5\times S^5$ Superstring}}} }
\bigskip\centerline{Nathan Berkovits\foot{e-mail: nberkovi@ift.unesp.br}}
\bigskip
\centerline{\it ICTP South American Institute for Fundamental Research}
\centerline{\it Instituto de F\'\i sica Te\'orica, UNESP - Univ. 
Estadual Paulista }
\centerline{\it Rua Dr. Bento T. Ferraz 271, 01140-070, S\~ao Paulo, SP, Brasil}
\bigskip

\vskip .3in

Using the pure spinor formalism for the superstring in an $AdS_5\times S^5$ background, a simple expression is found for half-BPS vertex operators.
At large radius, these vertex operators reduce to the usual supergravity vertex operators
in a flat background. And at small radius, there is a natural conjecture for generalizing these vertex operators to non-BPS states.

\vskip .3in

\Date {April 2019}
\newsec{Introduction}

Although the computation of superstring scattering amplitudes in an $AdS_5\times S^5$
background is complicated by the nonlinear form of the worldsheet action, the presence of
maximal supersymmetry and the duality with d=4 {\cal N}=4 super-Yang-Mills gives reasons to be optimistic that progress will be made. Since the RNS formalism can only be used to describe infinitesimal Ramond-Ramond backgrounds \MinahanFH\ChoNFN, one needs to use either the Green-Schwarz or pure spinor formalisms to fully describe $AdS_5\times S^5$.
The Green-Schwarz light-cone formalism is convenient for computing the
physical spectrum of ``long" strings \FrolovAV, but amplitude computations using this formalism are complicated even in a flat background.

The pure spinor formalism in an $AdS_5\times S^5$ background has the advantage over the Green-Schwarz formalism of allowing manifestly $PSU(2,2|4)$-covariant
quantization \MazzucatoJaT. Although less studied, this formalism was used to derive the quantum structure of the infinite set of nonlocal
conserved currents in \BerkovitsXU\ and to compute the physical spectrum of ``long" strings in \AisakaUD. And in a flat background, the pure spinor formalism has been used for computing multiloop superstring amplitudes \GomezSLA\
that have not yet been computed using either the RNS or Green-Schwarz formalisms.

To generalize these amplitude computations to an $AdS_5\times S^5$ background, the first
step is to explicitly construct the superstring vertex operators for half-BPS states. Although the behavior of half-BPS vertex operators near the $AdS_5\times S^5$ boundary was computed in \BerkovitsPS, the complete BRST-invariant vertex operator was only previously known for some special states \MikhailovAF\ such as the moduli for the $AdS$ radius \BerkovitsGA\ and for the $\b$-deformation \BedoyaQZ. 

In this paper, simple expressions will be obtained for general half-BPS vertex operators
in an $AdS_5\times S^5$ background using the pure spinor formalism. These expressions will be manifestly BRST-invariant and will closely resemble the
vertex operators for Type IIB supergravity states in a flat background. Hopefully, these
simple expressions for vertex operators will soon be used for computing superstring scattering amplitudes
in an $AdS_5\times S^5$ background.

In section 2, the BRST-invariant vertex operator for Type IIB supergravity states in a flat
background will be constructed in terms of the chiral supergravity superfield whose
lowest components are the dilaton and axion. In section 3, this vertex operator will
be expressed in a simple form using picture-changing operators. And in section 4,
this simple expression for the Type IIB supergravity vertex operator in a flat background
will be generalized to half-BPS vertex operators in an $AdS_5\times S^5$
background. Finally, section 6 will discuss the recent conjecture of \BerkovitsULM\ for generalizing this
construction to non-BPS states in an $AdS_5\times S^5$ background at small radius.

\newsec{Supergravity Vertex Operators}

In any Type IIB supergravity background, the massless closed superstring vertex operator in
unintegrated form in the pure spinor formalism is\BerkovitsYR
\eqn\any{V = \l_L^\a \l_R^\b A_{\a\b} (x,\t_L,\t_R)}
where $A_{\a\b}$ are bispinor superfields depending on the N=2B d=10 superspace variables $(x^m,
\t_L^\a,\t_R^\a)$, $\a =1$ to 16 are Majorana-Weyl spinor indices, and $\l_L^\a$ and $\l_R^\a$ are left and right-moving
pure spinor variables satisfying $\l_L\g^m\l_L = \l_R\g^m \l_R=0$ for $m=0$ to 9. The onshell equations of motion and gauge invariances are implied by $QV =0$ and $\d V = Q \O$ where
\eqn\brstcurved{Q = \l_L^\a \nabla_{L\a} + \l_R^\a \nabla_{R\a}}
 and $\nabla_{L\a}$ and $\N_{R\a}$ are the 32 fermionic covariant derivatives in the supergravity background.
 These equations of motion and gauge invariances imply that $A_{\a\b}$ satisfies
 \eqn\satisfy{\g_{abcde}^{\a\g} \nabla_{L\a} A_{\g\b} = \g_{abcde}^{\a\b} \N_{R\a} A_{\g\b} =0,\quad
 \d A_{\a\b} = \nabla_{L\a} \O_{R\b} + \N_{R\b} \O_{L\a},}
 where $\O_{L\a}$ and $\O_{R\a}$ satisfy $\g_{abcde}^{\a\b} \nabla_{L\a}\O_{L\b} =
 \g_{abcde}^{\a\b} \N_{R\a} \O_{R\b} = 0$.

 \subsec{Flat background}
 
 To construct solutions to \satisfy\ in a flat background, it is convenient to choose a reference frame where the momentum is only in the $k_+ = k_0 + k_9$ direction so that the covariant fermionic derivatives reduce to
 \eqn\flatN{\N_{La} \equiv (\g^- \N_L)_a = {\p\over{\p\t_L^a}} + \t_{La} \p_+, 
 \quad \N_{L\ad} \equiv (\g^+\N_L)_\ad = {\p\over{\p\tb_L^\ad}} , }
 $$\N_{Ra} \equiv (\g^- \N_R)_a = {\p\over{\p\t_R^a}} + \t_{Ra} \p_+, 
 \quad \N_{R\ad} \equiv (\g^+\N_R)_\ad = {\p\over{\p\tb_R^\ad}} ,$$
 where $a,\ad$ are SO(8) chiral and antichiral spinor indices and 
 \eqn\ttt{\t_{La} = (\g^+\t_L)_a, \quad\tb_{L\ad} = (\g^- \t_L)_\ad,\quad  \t_{Ra} = (\g^+\t_R)_a,\quad \tb_{R\ad} = (\g^- \t_R)_\ad.} 
 
 Since $k_+$ is nonzero, \satisfy\ implies one can gauge-fix
 $A_{a \bd} = A_{a b} = A_{\ad b} =0$, so that
 \eqn\lcgauge{ V = \lb_L^\ad \lb_R^\bd A_{\ad \bd}(x,\t_L,\t_R)}
 where $\l_L^a = (\g^+\l_L)^a$, $\lb_L^\ad = (\g^- \l_L)^\ad$,   $\l_R^a = (\g^+\l_R)^a$, $\lb_R^\ad = (\g^- \l_R)^\ad$. 
 In the gauge of \lcgauge, $QV =0$ together with $\lb_L^\ad \l_L^a \s^j_{a\ad} = \lb_R^\ad \l_R^a \s^j_{a\ad} =0$ implies that
  \eqn\pluseq{ {\p\over{\p\tb_L^\ad}} A_{\bd\cd} =  {\p\over{\p\tb_R^\ad}} A_{\bd\cd}=0, \quad \N_{L a} A_{\bd \cd}  ={1\over 8}\s^j_{a\bd} \s_j^{c \dd} \N_{Lc} A_{\dd \cd}, \quad
 \N_{R a} A_{\bd \cd} ={1\over 8} \s^j_{a\cd} \s_j^{c \dd} \N_{Rc} A_{\bd \dd} }
 where $\s^j_{a\ad}$ are the SO(8) Pauli matrices. 
 
One method of solving \pluseq\ is to take the left-right product of the open superstring solutions of \jusinskas, but it will be useful to describe another method which can be easily generalized to the $AdS_5\times S^5$ background. This method is based on the SO(8) chiral superfield $\Phi$ satisfying $\N_-^a \Phi=0$ where $\N_\pm^a \equiv \N_{L}^a \pm i \N_{R}^a$ is a linear combination of the left and right-moving
fermionic derivatives. In terms of $(x^+, \t_L^a, \t_R^a)$, 
\eqn\phidef{\Phi (x^+, \t_L^a, \t_R^a)= e^{ i k_+ (x^+ + i\t_L^a\t_R^a)} f(\t_-)}
where $\t_-^a = \t_L^a - i \t_R^a$. The superfield $\Phi$ will be defined to satisfy the reality condition
$(\N_+)^4_{abcd} \Phi ={1\over {24}}\e_{abcdefgh}  (\N_-)_{efgh}^4 \overline\Phi$, and the $2^8$ components of $\Phi$ describe the Type IIB supergravity multiplet where, at zeroth order in $\t_-$, the real part of $\Phi$ is the Type IIB dilaton and the imaginary part of $\Phi$ is the Type IIB axion.

To construct the vertex operator of \lcgauge\ for this multiplet, first consider the vertex operator
\eqn\startv{ V_0 = \lb_L^\ad \lb_R^\ad  \Phi .}
Using the relation $\l_L^a \lb_L^\ad =-{1\over 4} (\s^{jk}\l_L)^a (\s_{jk}\lb_L)^\ad$ and $\l_R^a \lb_R^\ad = -{1\over 4} (\s^{jk}\l_R)^a (\s_{jk}\lb_R)^\ad$,  
one finds that 
\eqn\relone{ Q V_0 = (\l_- \N_+ + \l_+ \N_-) (\lb_L \lb_R) \Phi = (\l_-\N_+)(\lb_L\lb_R)\Phi  
	 = -{1\over 4}(\l_+\s^{jk} \N_+) (\lb_L\s_{jk}\lb_R) \Phi}
 where $\l_\pm^a = \l_L^a\pm i \l_R^a$. Now consider the vertex operator
 \eqn\secondv{ V_1 = {1\over {32i k_+}} (\lb_L \s_{jk} \lb_R) (\N_+ \s^{jk} \N_+) \Phi .}
 Since $\{\N_-, \N_+\} =4 \p_+$, \relone\ implies that
 $QV_0 = - (\l_+ \N_-) V_1$. Furthermore, a similar argument implies that 
 $(\l_- \N_+) V_1 = -(\l_+ \N_-) V_2$ where
  \eqn\threev{ V_2 = -{1\over{2048 k_+^2}} (\lb_L \s_{jklm} \lb_R) (\N_+ \s^{jk} \N_+) (\N_+ \s^{lm} \N_+)\Phi .}
  Continuing this argument, one finds that $QV =0$ where
  $V = V_0 + V_1 + V_2 + V_3 + V_4$ and
 \eqn\allv{ V_n =  {1\over{n! (32ik_+)^n}}(\lb_L \s_{j_1 k_1 ... j_n k_n} \lb_R) (\N_+ \s^{j_1 k_1} \N_+) ... (\N_+ \s^{j_n k_n} \N_+)  \Phi .}
 Note that $(\l_-\N_+) V_4=0$ since $(\N_+)^9 \Phi =0$. 
 
 So the BRST-invariant vertex operator with momentum $k_+$ in this gauge is
 \eqn\final{V = \lb_L^\ad \lb_R^\bd e^{ik_+ x^+} A_{\ad\bd}(\t_L,\t_R) =V_0+V_1+V_2+V_3+V_4 , }
 and one can easily verify that at $\t_L^a =\t_R^a =0$, $A_{\ad\bd}$ is the bispinor Ramond-Ramond field in light-cone gauge
 \eqn\bispinorrr{A_{\ad\bd} = \d_{\ad\bd} a + \s^{jk}_{\ad\bd} a_{jk} + \s^{jklm}_{\ad\bd} a_{jklm}.}
 It will be useful to note that one would end up with the same expression of \final\ for $V$ if one had instead started with the superfield $\Phi_{1234}$ which is annihilated by $\nabla_-^a \equiv \N_L^a -i (\s_{1234}\N_R)^a$. In this case, $V_0 = (\lb_L\s_{1234}\lb_R) \Phi_{1234}$ and
 \eqn\allvtwo{ V_n =   {1\over{n! (32ik_+)^n}}(\lb_L \s_{j_1 k_1 ... j_n k_n} \s_{1234}\lb_R) (\N_+ \s^{j_1 k_1} \N_+) ... (\N_+ \s^{j_n k_n}\N_+)  \Phi_{1234} }
 where $\N_+^a \equiv \N_L^a +i (\s_{1234}\N_R)^a$.

 
 \newsec{Picture-Changing}
  
 To generalize this construction to an $AdS_5\times S^5$ background, it will be useful to first consider the vertex operator $V$ for the lowest component of $\Phi_{1234}$ in
 \allvtwo, i.e. $\Phi_{1234}= \exp (ik_+ \hat x^+)$ where $\hat x^+ \equiv x^+ + i\t_L \s_{1234}\t_R$ . Although this vertex operator of \allvtwo\ has various terms $V_0 ... V_4$ with different powers of $\t_+^a=
 \t_L^a + i (\s_{1234}\t_R)^a$, it can be reduced to just one term by writing it in a different ``picture" as 
 \eqn\plv{V_{-1} = P V  =  (\lb_L\s_{1234}\lb_R) e^{ik_+ \hat x^+} \prod_{a=1}^8 \t_+^a \d (\l^a_+)}
 where $P$ is the ``picture-lowering" operator 
 \eqn\Ddef{P =  \prod_{a=1}^8 \t_+^a \d (\l^a_+)}
 and  $\l_+^a = \l_L^a + i(\s_{1234}\l_R)^a$. Note that the 8 $\l_+^a$'s in $P$ are all independent so that $\prod_{a=1}^8 \d (\l^a_+)$ is well-defined.
 Also note that $P$ is BRST-invariant and is super-Poincar\'e invariant up to a BRST-trivial quantity. For example, under the supersymmetry transformation generated by $q_1$, 
 \eqn\susyD{ q_1 P = \d (\l^1_+)\prod_{a=2}^8 \t_+^a \d(\l^a_+) = Q [- \t_+^1 \d'(\l^1_+) \prod_{a=2}^8 \t_+^a \d(\l^a_+)].}

 The original vertex operator $V$ of \final\ is related to $V_{-1}$ of \plv\ by picture-raising as
 $V = C V_{-1}$
 where 
 \eqn\pro{C = \prod_{a=1}^8 Q(\xi_a) }
  is the picture-raising operator and $Q(\xi_a)$ is a formal expression whose action on $V_{-1}$ is defined through the following procedure: Using the notation of Friedan-Martinec-Shenker for 
  picture-changing operators, $\d (\gamma) = e^{-\phi}$ and $\xi \d(\g) = \xi e^{-\phi} = {1\over\g}$ where $(\g,\b)$ are chiral bosons which have been fermionized as 
$\g = \eta e^\phi$ and $\beta = \p\xi e^{-\phi}$. Although $\l_+^a$ and its conjugate $w_a^+$ are not chiral bosons, one can formally define
\eqn\fms{\l_+^a = \eta^a e^{\phi_a}, \quad w_a^+ = \p\xi_a e^{-\phi_a}}
so that 
\eqn\converx{\xi_a \d(\l_+^a) = \xi_a e^{-\phi_a} = {1\over {\l_+^a}}.}

Using this definition, $C V_{-1}$ can be computed by using \converx\ to convert the factors of $\d(\l_+^a)$ in $V_{-1}$ into factors of ${1\over {\l_+^a}}$. Furthermore, the BRST invariance of $V_{-1}$ guarantees that $C V_{-1}$ has no poles when $\l_+^a=0$  and can be expressed in the form of \any\ as $V = \l_L^\a \l_R^\b A_{\a\b}(x,\t_L, \t_R)$. 
To see why, note that $Q( F \delta (\l_+^ a))=0$ implies that $Q(F)$ is proportional to $\l_+^a$.  
So $Q({F\over{\l_+^a}})$ has no poles when $\l_+^ a=0$. Also note that if $F$ has (left,right)-moving ghost number equal to $(g_L, g_R)$, then $Q({F\over{\l_+^a}})$ also has (left,right) ghost number $(g_L, g_R)$. This is easy to see since terms in $QF$ must either carry ghost number $(g_L+1, g_R)$ or $(g_L, g_R+1)$. So $QF = E \l_+^a$ for some $E$ implies that $E$ must carry ghost number $(g_L, g_R)$.

One can explicitly compute $C V_{-1}$ for the vertex operator of \plv\ as
  \eqn\qeta{C V_{-1} =\prod_{b=2}^8 Q(\xi_b) ~Q (\xi_1) V_{-1} = \prod_{b=2}^8 Q(\xi_b) ~Q (\xi_1 V_{-1}) }
  $$=- \prod_{b=2}^8 Q(\xi_b)~ Q ({{\t_+^1}\over{\l_+^1}}\prod_{a=2}^8 \t_+^a \d(\l_+^a ) (\lb_L\s_{1234} \lb_R) e^{ik_+ \hat x^+})$$
  $$
   =- \prod_{b=2}^8 Q(\xi_b)~\prod_{a=2}^8 \t_+^a \d(\l_+^a)(\lb_L\s_{1234} \lb_R)e^{ik_+ \hat x^+}$$
  $$ =-  \prod_{b=3}^8 Q(\xi_b) ~Q(\xi_2\prod_{a=2}^8 \t_+^a \d(\l_+^a)  (\lb_L\s_{1234} \lb_R) e^{ik_+\hat x^+})
  $$
  $$
  = \prod_{b=3}^8 Q(\xi_b) ~Q({{\t_+^2}\over{\l_+^2}} \prod_{a=3}^8 \t_+^a \d(\l_+^a)(\lb_L\s_{1234} \lb_R) e^{ik_+ \hat x^+})$$
  $$ = \prod_{b=3}^8 Q(\xi_b) ~ (1 +2i k_+ (\l_-^1 \t_+^1) {{\t_+^2}\over{\l_+^2}} )\prod_{a=3}^8 \t_+^a \d(\l_+^a) (\lb_L\s_{1234} \lb_R) e^{ik_+ \hat x^+} $$
  $$= \prod_{b=3}^8 Q(\xi_b) ~ ((\lb_L\s_{1234} \lb_R)+{i\over 2} k_+ \t_+^1\t_+^2 \s^{jk}_{12} (\lb_L \s^{jk}\s_{1234} \lb_R) )\prod_{a=3}^8 \t_+^a \d(\l_+^a)   e^{ik_+ \hat x^+}$$
  where we have used that $\l_-^1 (\lb_L\s_{1234}\lb_R) ={1\over 4} (\s^{jk} \l_+)^1 (\lb_L \s^{jk}\s_{1234} \lb_R)$. Continuing with this procedure of converting $\xi_a \d (\l_+^a)$ into $(\l_+^a)^{-1}$ to compute the product with $Q(\xi_a)$, it is expected that
  $C V_{-1}$ will reproduce $V$ of \final. 
  
\newsec{$AdS_5\times S^5$ Vertex Operators}

\subsec{ Parameterization of $AdS_5\times S^5$}

To generalize this construction for half-BPS states in an $AdS_5\times S^5$ background, parameterize $AdS_5\times S^5$ using the
supercoset $g\in {{PSU(2,2|4)}\over{SO(4,1)\times SO(5)}}$ as
\eqn\superc{g (\t, X, Y) = F(\t) G(X) H(Y)}
where $F(\t) = \exp (\t^J_R q_J^R + \t^R_J q^J_R)$ is a fermionic ${{PSU(2,2|4)}\over{SO(4,2)\times SO(6)}}$ coset, $(q^R_J, q^J_R)$ are the 32
fermionic generators of $PSU(2,2|4)$, $R=1$ to 4 are $SO(4,2)$ spinor indices, $J=1$ to 4 are $SO(6)$ spinor indices, $G(X)$ is an
${{SO(4,2)}\over{SO(4,1)}}$ coset for $AdS_5$ and $H(Y)$ is an ${{SO(6)}\over{SO(5)}}$ coset for $S^5$. Under global $PSU(2,2|4)$ transformations, $\d g = \Sigma g$ where $\Sigma \in PSU(2,2|4)$, and under BRST transformations, 
\eqn\brstg{\d g = g [(\l_L +i \l_R)^{\tilde J}_{\tilde R} q^{R}_{J} + (\l_L -i \l_R)_{\tilde J}^{\tilde R} q_{R}^{J}]}
where $\tilde R=1$ to 4 is an $SO(4,1)$ spinor index, $\tilde J=1$ to 4 is an $SO(5)$ spinor index, and $(\l_L)^{\tilde R}_{\tilde J}$ and $(\l_R)^{\tilde R}_{\tilde J}$ are the left and right-moving pure spinors. Note that $SO(4,1)$ and $SO(5)$ spinor indices can
be raised and lowered using the matrices $\s_6^{\tilde R \tilde S}$ and $\s_6^{\tilde J \tilde K}$ which commute with $SO(4,1)$ and $SO(5)$ rotations.

The cosets $G(X)$ and $H(Y)$ are defined up to local $SO(4,1)\times SO(5)$ gauge transformations parameterized by $\Omega\in SO(4,1)$ and $\hat\Omega\in SO(5)$ as
\eqn\cosets{G(X) \sim G(X) \Omega, \quad H(Y) \sim H(Y) \hat\Omega}
where the left and right-moving pure spinors $\l_L$ and $\l_R$ transform as $SO(4,1)\times SO(5)$ spinors. More explicitly, 
$G^R_{\tilde R}$ and $H^J_{\tilde J}$ are $4\times 4$ matrices which transform under the gauge transformations as
\eqn\cosets{G^R_{\tilde R} \to G^R_{\tilde S} \Omega^{\tilde S}_{\tilde R}, \quad H^J_{\tilde J} \to H^J_{\tilde K} {\hat\Omega}^{\tilde K}_{\tilde J}, }
$$(\l_L)^{\tilde R}_{\tilde J} \to (\l_L)^{\tilde S}_{\tilde K}  \Omega^{\tilde R}_{\tilde S} {\hat\Omega}^{\tilde K}_{\tilde J}, \quad 
(\l_R)^{\tilde R}_{\tilde J} \to (\l_R)^{\tilde S}_{\tilde K}  \Omega^{\tilde R}_{\tilde S} {\hat\Omega}^{\tilde K}_{\tilde J},$$
and the $AdS_5$ coordinate $X^{RS}= -X^{SR}$ and
$S^5$ coordinate $Y^{JK} = - Y^{KJ}$ are defined in terms of $G^R_{\tilde R}$ and $H^J_{\tilde J}$ by
\eqn\xy{X^{RS} = G^R_{\tilde R} \s_6^{\tilde R \tilde S} G^S_{\tilde S}, \quad Y^{JK} = H^J_{\tilde J} \s_6^{\tilde J \tilde K} H^K_{\tilde K}.}
Defining $X_{RS} = \half \e_{RSTU} X^{TU}$ and $Y_{JK} = \half \e_{JKLM} Y^{LM}$, \xy\ implies $X^{RS} X_{RS} =4$ and
$Y^{JK} Y_{JK} = 4$. 

\subsec{ Half-BPS vertex operator}

To construct the vertex operator for a half-BPS state in an $AdS_5\times S^5$ background, consider the state dual to the super-Yang-Mills
gauge-invariant operator 
\eqn\sym{Tr [(y_0^{JK} \Phi_{JK}(x))^n]}
 where $\Phi_{JK}(x)$ are the six scalars located at the position $x^m$ on the $AdS_5$ boundary and $y_0^{JK}$ is a fixed null six-vector satisfying $\e_{JKLM} y_0^{JK} y_0^{LM} =0$. It will be convenient to define the null six-vector
\eqn\nullsix{x_0^{RS} = (\e^{AB}, x^m \s_m^{A\dot A}, (x^m x_m) \e^{\dot A\dot B})}
where $R=(A,\dot A)$ with $A,\dot A=1$ to 2. $x_0^{RS}$ transforms covariantly under $SO(4,2)$ conformal transformations of the $AdS_5$ boundary and satisfies $\e_{RSTU} x_0^{RS} x_0^{TU} =0$.

The choice of $y_0^{JK}$ breaks $SO(6)$ $R$-symmetry to $U(1)\times SO(4)$, and $J$ will be defined to be the charge with respect to this $U(1)$. Similarly, the choice of
$x_0^{RS}$ breaks $SO(4,2)$ conformal symmetry to $SO(1,1)\times SO(3,1)$, and $\Delta$ will be defined to be the charge with respect to the $SO(1,1)$. The half-BPS state of \sym\ carries $J= n$ and $\Delta = n$ and is preserved by the 24 spacetime supersymmetries which carry $J-\Delta\geq 0$.

 In analogy with the construction of the vertex operator of $V_{-1}$ in a flat background, it will now be argued that the BRST-invariant vertex operator for the state
\sym\ is
\eqn\vads{V_{-1} = (\l_L)^{\tilde J}_{\tilde R} (\l_R)^{\tilde R}_{\tilde J} ~ P ~ ({{Y\cdot  y_0}\over{X\cdot  x_0}} )^n}
where the picture-lowering operator $P$ is defined as
\eqn\ploads{P =\prod_{a=1}^8 \t_+^a \d(Q(\t_+^a)) }
and $\t_+^a$ are the 8 $\t$'s which carry charge $J-\Delta=1$. In terms of
$x_0^{RS}$ and $y_0^{JK}$, 
\eqn\expt{\t_+^a = 
[(x_0)^{RS} (y_0)_{JK}\t^K_S ,\,\, (x_0)_{RS}(y_0)^{JK} \t_K^S ]}
where only 8 of the 32 components of $(x_0)^{RS} (y_0)_{JK} \t^K_S$ and
$(x_0)_{RS}  (y_0)^{JK}\t_K^S$ are independent since $(x_0)^{RS} (x_0)_{ST} = (y_0)^{JK} (y_0)_{KL} =0$.


To show that $V_{-1}$ of \vads\ carries the same charges and is invariant under the same 24 supersymmetries as \sym, note that $Y\cdot y_0$ carries $J=1$ and 
$X\cdot  x_0$ carries $\Delta =-1$ so that $V_{-1}$ carries $J=\Delta = n$. Furthermore, both $Y\cdot y_0$ and $X\cdot x_0$ are invariant
under the 8 supersymmetries with $J -\Delta = 1$. And under the 16 supersymmetries with $J - \Delta = 0$,
${{Y\cdot y_0}\over{X\cdot x_0}}$ transforms into terms which contain at least one $\t$ with $J -\Delta =1$. However, all 8 $\theta$'s with
$J -\Delta = 1$ are contained in the picture-lowering operator $P$ of \ploads. So $V_{-1}$ is invariant under all 24 supersymmetries which carry
$J - \Delta\geq 0$. 

Similarly, under the BRST transformation of \brstg, ${{Y\cdot y_0}\over{X\cdot  x_0}}$ transforms into terms containing products of
$Q(\t)$ with $\t$'s where either $Q(\t)$ carries $J - \Delta =1$ or at least one of the $\t$'s carries $J-\Delta =1$. In both cases, the BRST transformation
is killed by $P= \prod_{a=1}^8 \t_+^a \d(Q(\t_+^a))$ of \ploads. And since $P$ and $(\l_L)^{\tilde J}_{\tilde R} (\l_R)^{\tilde R}_{\tilde J} $ are also BRST-invariant, it has been shown that $V_{-1}$ of \vads\ is BRST-invariant. 

\subsec{Explicit example}

For example, consider the state corresponding to $Tr[(\Phi_{12}(0))^n]$ which carries 
$\Delta =J =n$ where $\Delta$ is the dilatation charge and $J$ is the $U(1)$ charge.
To simplify the vertex operator, parameterize the supercoset $g\in {{PSU(2,2|4)}\over{SO(4,1)\times SO(5)}}$ as
\eqn\paramn{g = \exp (\t_{-} q_{+} + \t_0 q_0 + x K + y R) \exp( \t_{+} q_{-})
\exp (z \Delta + w J)}
where $(q_{+}, q_0, q_{-})$ are the $(8,16,8)$ fermionic isometries with $(+1, 0, -1)$ charge with respect to $J-\Delta$, and $K$ and $R$ are the four conformal boosts and four $R$-symmetries with charge $J-\Delta =1$. Since the vertex operator $V$ is annihilated by $(q_{+}, q_0, K, R, \Delta -J)$, the parameterization of \paramn\ implies that $V$ is independent of $(\t_{-}, \t_0, x, y, w+z)$ and only depends on $(\t_{+}, z-w)$ and the pure spinor ghosts.

Using the picture-lowering operator
$P = \prod_{a=1}^8 \t_{+}^a \d (Q(\t_+^a))$, the vertex operator
of \vads\ is
\eqn\vadse{V_{-1} =   
(\lb_L\s_{1234}\lb_R) e^{n(w-z)} \prod_{a=1}^8 \t_{+}^a \d (\l_+^a) }
where 
$(\l_+^a, \l_-^a, \lb_L^{\dot a}, \lb_R^{\dot a})$ for $a,\dot a=1$ to 8 are defined by
$$ \l_+^a \equiv [ e^{\half (z-w)} (\l_L+i\l_R)_{\tilde R}^{\tilde J} \,
{\rm for} \,\tilde J=1,2, \tilde R=1,2;\quad
e^{\half (z-w)} (\l_L-i\l_R)_{\tilde R}^{\tilde J} \,
{\rm for} \,\tilde J=3,4, \tilde R=3,4 ]$$
$$ \l_-^a \equiv [ e^{\half (w-z)} (\l_L+i\l_R)_{\tilde R}^{\tilde J} \,
{\rm for} \,\tilde J=3,4, \tilde R=3,4;\quad
e^{\half (w-z)} (\l_L-i\l_R)_{\tilde R}^{\tilde J} \,
{\rm for} \,\tilde J=1,2, \tilde R=1,2 ]$$
$$ \lb_L^{\dot a} \equiv [ (\l_L)_{\tilde R}^{\tilde J} \,
{\rm for} \,\tilde J=3,4, \tilde R=1,2;\quad  (\l_L)_{\tilde R}^{\tilde J} \,
{\rm for} \,\tilde J=1,2, \tilde R=3,4]$$
\eqn\lambdadefs{ \lb_R^{\dot a} \equiv [ (\l_R)_{\tilde R}^{\tilde J} \,
{\rm for} \,\tilde J=3,4, \tilde R=1,2;\quad  (\l_R)_{\tilde R}^{\tilde J} \,
{\rm for} \,\tilde J=1,2, \tilde R=3,4]}
and we have used that
$ (\l_L)^{\tilde J}_{\tilde R} (\l_R)^{\tilde R}_{\tilde J} = \lb_L^{\dot a} (\s_{1234})_{\dot a \dot b} \lb_R^{\dot b}$
when $\l_+^a=0$.

In the large radius limit where the $AdS_5\times S^5$ background approaches flat space, one can easily verify that $V_{-1}$ of \vadse\ approaches the flat space vertex operator $V_{-1}$ of \plv\ where
$k_+ =n$ and $i x^+$ is identified with $w-z$.
And the vertex operator for all other half-BPS states in an $AdS_5\times S^5$ background are obtained from \vadse\ by acting with the appropriate
$PSU(2,2|4)$ transformations, and reduce in the flat space limit to the vertex operators of other supergravity states in the muitiplet of \plv. 

Finally, one can relate $V_{-1}$ of \vadse\ to the supergravity vertex operator $V= \l_L^\a \l_R^\b A_{\a\b}(x,\t)$ of \any\ by defining 
\eqn\cvadas{V = C V^{-1}}
where 
$C = \prod_{a=1}^8 Q(\xi_a)$ and the 8 $\l_+^a$'s of \lambdadefs\ have been fermionized as in \fms. Using the same procedure as in \qeta, this construction will produce an $AdS_5\times S^5$ vertex operator of the form  
$V= \l_L^\a \l_R^\b A_{\a\b}(\t_+, z-w)$ where, as in a flat background, the potential
poles coming from $\xi^a \delta(\l_+^a) = {1\over {\l_+^a}}$ are absent because of the BRST invariance of $V_{-1}$.   

\newsec{Summary}

In this paper, a simple BRST-invariant vertex operator was constructed for half-BPS states in an $AdS_5\times S^5$ background. One possible application of this paper is to use these vertex operators to compute scattering amplitudes. Much is known about scattering amplitudes of half-BPS states in $AdS_5\times S^5$, and it would be very interesting to show how to compute these amplitudes using superstring vertex operators even for the simplest 3-point amplitude.

Another possible application of this paper is to construct $AdS_5\times S^5$ vertex operators for non-BPS states. As discussed in \BerkovitsULM, the half-BPS vertex operator can be expressed as
\eqn\vnonbps{ V = (\l_L)^{\tilde R}_{\tilde J} (\l_R)^{\tilde J}_{\tilde R}  (C ~ P  {{Y\cdot  y_0}\over{X\cdot  x_0}})^n}
if one adds $(n-1)$ picture-raising operators $C$ and $(n-1)$ picture-lowering operators $P$ to $V= C V_{-1}$ of \cvadas. Since all states at zero 't Hooft coupling
can be described as ``spin chains" constructed from $n$ super-Yang-Mills fields, it is natural to express the half-BPS vertex operator of \vnonbps\ as
\eqn\vnonbpsatwo{ V = (\l_L)^{\tilde R}_{\tilde J} (\l_R)^{\tilde J}_{\tilde R}  ~C ~ E ~C ~ E ~ ... ~ C ~ E}
where $E \equiv P {{Y\cdot y_0}\over{X\cdot  x_0}}$ corresponds to the Yang-Mills field
$y_0^{JK} \phi_{JK} (x_0)$ on the spin chain. Therefore,
 a natural conjecture for general non-BPS vertex operators is
\eqn\vnonbpstwo{ V = (\l_L)^{\tilde R}_{\tilde J} (\l_R)^{\tilde J}_{\tilde R}  ~: C ~ E_1 ~C ~ E_2 ~ ... ~ C ~ E_n :}
where $E_1 ... E_n$ describe $n$ different super-Yang-Mills fields on the spin chain and are obtained from $ P  {{Y\cdot y_0}\over{X\cdot  x_0}}$ by performing
the appropriate $PSU(2,2|4)$ transformation. Since $E$ and $C$ are independently BRST-invariant, the vertex operator of \vnonbpstwo\ is BRST-invariant where $:~:$ denotes a normal-ordering prescription which is defined to be invariant under cyclic permuations of the
$E$'s. It would be very interesting to find evidence for this conjecture by using the topological description of \BerkovitsULM\ to study the $AdS_5\times S^5$ superstring
at small radius. 

\vskip15pt
{\bf Acknowledgements:}
I would
like to thank Thales Agricola, Thiago Fleury, Juan Maldacena, Andrei Mikhailov and Pedro Vieira
for useful discussions, and
CNPq grant 300256/94-9
and FAPESP grants 2016/01343-7 and 2014/18634-9 for partial financial support.

\listrefs
\end




\listrefs
\end